\def\Godel{G\"odel }
\def\godel{G\"odel}
\def\Godels{G\"odel's }

\def\nat{{\cal N}}
\def\exp{\hbox{exp}}

\def\citeclay{[1]}
\def\Citeclay{[1] }

\hsize= 13.5 cm\hoffset = 2cm

\def\section#1{\vskip 1cm\centerline{{\bf #1}}\vskip.2cm}
\def\proclaim#1#2{\vskip0cm\noindent {\bf #1 }{\sl #2}}
\def\proof{\vskip0cm\noindent {\bf Proof.} }

\def\endproof{\hbox{$\hskip.2cm{\vrule height5pt width5pt depth 0pt}$}}

\centerline{{\bf Constructive solvability and the P versus NP problem}}

\vskip.5cm

\centerline{Arne Hole}
\centerline{University of Oslo}
\centerline{Arne.Hole@ils.uio.no}

\vskip.65cm

\centerline{{\bf Abstract}}

\vskip.3cm

\noindent {\it The relation between the computational complexity class NP and other complexity classes is addressed in the context of provability and limitations on the possibility of finding sound axioms for formal theories. We construct a family D of decision problems and show that under a certain finiteness condition, D contains a problem which is in NP. Further, it is shown that if the term ``constructible theory'' is defined in a way satisfying a specific natural condition, then no constructible and sound theory verifies a solution algorithm for any of the problems in D. Arguably, this solves the P versus NP problem under a constructive interpretation. The relation to classical proofs of NP $\subseteq$ EXPTIME is discussed. These proofs tacitly use an assumption which may fail for problems in D.}

\section{1. Introduction}

\noindent For general information on the P versus NP problem, see \citeclay. In short form,
the problem is often described verbally in a way similar to the following: 
{\it If we can recognize a correct solution to a given decision problem effectively, then is it always the case that the problem may also be solved effectively?}
Or using some slightly more technical words:

\vskip.15cm

\item{}{\it If we can verify a correct solution to a given decision problem in polynomial time, then is it always the case that the problem may also be solved in polynomial time?}

\vskip.15cm

\noindent The P versus NP problem is asking for a proof answering  a precise version \Citeclay of this question. However, given the extremely wide scope of the problem, 
with algorithms (Turing machines) modelling the general idea of a ``method'' through the Church-Turing thesis, we 
need to be precise about what is meant by a ``proof'' in this context.

In the modern mathematical interpretation of the word, a proof is an
argument which may be translated into a proof in some formal theory $T$. Moreover, if we want to ``establish'' a mathematical result concerning, say, natural numbers, then $T$ must be 
{\it arithmetically sound\/} in the sense that theorems of $T$ which can be translated into statements about 
the set $\nat$ of natural numbers, are considered to be true 
under that interpretation. In this paper, the word ``sound'' means arithmetically sound.

There is an obvious link between soundness of formal theories and verifiability of algorithms. Having constructed an algorithm for solving a problem without being able to prove that the algorithm actually solves the problem, is of little use. We need to verify the algorithm as a solution by a proof in a theory considered to be sound. 

The deep logical problem here is the element of subjectivity introduced by the phrase ``considered to be'', which in all meaningful applications is indispensable. Proof in a theory whose soundness is in doubt, is insufficient. Also, provability in a theory which we can only describe indirectly, is not sufficient for establishing a mathematical result. For instance, the theory obtained from ZF by adding either $\hbox{P}=\hbox{NP}$ or 
$\hbox{P}\neq\hbox{NP}$ as an axiom, depending on which one of them is ``true'', clearly is a meaningsless choice in the P versus NP context. We need a proof in a theory that we {\it actually have constructed\/}, 
with axioms which are all considered as {\it a priori\/} self-evident for the intended models, so that we have a complete proof available by direct reference. In other words, we need an explicitly constructed, sound theory, with ``sound'' meaning ``arithmetically sound''. In this paper, we assume that Peano Arithmetic (PA) is a sound theory.

The constructibility requirement is naturally reflected in the informal descriptions of the P versus NP problem given here, through phrases such as ``If we can'', which clearly indicates constructibility. Of course, the widespread interest in the P versus NP problem, and more generally the theory of computational complexity classes such as P and NP, stems from the prospect of applications where one explicitly constructs algorithms to solve some problem, and seeks to verify the algorithms as solutions in some explicitly constructed theory assumed to be sound. Limiting the discussion to algorithms satis\-fying this condition, we obtain what one may refer to as the {\it constructive\/} interpretation computational complexity classes. Throughout the paper, we assume this interpretation.

In Section 2, we define a set $G$ which represents, in a sense, all mathematical theories in which proofs can be ``formalized'', regardless of logic and language. We then define a family $D=\lbrace D_k\mid k\in\nat\rbrace$ of decision problems
related to $G$.
In Section 4, we state a condition CC on theories 
represented in $G$ which are to be labeled {\it constructible\/}, and a finiteness condition FC concerning $D$. We then obtain the following result (Theorem 5):

\vskip.3cm

\item{}{\it Assume CC and FC. Then there exists $k\in\nat$ such that (i) $D_k\in\hbox{NP}$, and
\item{}(ii) No sound, constructible theory represented in $G$ proves $D_k\in\hbox{P}$.}

\vskip.3cm

\noindent In Section 5, we argue that this result solves the P versus NP problem under the constructive interpretation of 
computational complexity classes. The conclusion is that 
$\hbox{NP}\not\subseteq\hbox{P}$. However, since formally speaking both CC and FC are taken for granted, and CC is not even a formalizable mathematical condition, such a solution may not be what some people have been hoping for. Concerning this, well known arguments can be given for the case that finding a proof of $\hbox{NP}\not\subseteq\hbox{P}$ in
one {\it fixed\/}, constructible and sound theory, is impossible {\it a priori\/}. Involvement of uformalizable condition
such as CC is then unavoidable. We will return to this point in Section 5. 

\section{2. Definitions}

\noindent In this paper, $L$ will denote the language  $L(\hbox{PA})$ of 
Peano Arithmetic (PA). When we speak of truth or falsity of $L$-formulas, we 
always refer to truth in the standard structure $\nat$ of 
natural numbers $\lbrace 0,1,2,3\ldots\rbrace$. If $n\in\nat$, then ${\bf n}$ denotes the
numerical term $SS...SO$ representing $n$, with $n$ copies of the successor relation symbol $S$.
In the following, basic definitions are given relative to 
a chosen \Godel numbering $\gamma=\lceil ...\rceil$ of $L$. Sequences of formulas, and in particular PA proofs, are 
assigned $\gamma$ numbers in the usual way. 
In the text, we use standard abbreviations of $L$ formulas. Throughout the paper, the word {\it algorithm\/} means {\it Turing machine\/}. 

Our approach is to consider mathematical theories $T$ for which provability of $L$-formulas may be expressed by some $L$-formula 
$$
A(x,y)
$$ 
with the two free variables $x$ and $y$, in the sense that
$$
\exists y A(x,y)\eqno(1)
$$
is true iff $x$ is the $\gamma$ number of an $L$-formula $B$, and $B$ is provable in $T$. Note that aside from the fact that the language of $T$ must contain the symbols needed for expressing the $L$-formula $B$, in this approach we make no assumptions on the language of $T$. In effect, we cover all recursively axiomatized theories proving sentences of $L$ ``up to language''. As an example, consider ZF. This theory can in effect derive all
theorems of PA, but they will be expressed differently, namely in the set-theoretic 
language $L(\hbox{ZF})$. However, we may construct from ZF an equally strong theory 
$\hbox{ZF'}$ by extending $L(\hbox{ZF})$ with the non-logical symbols of $L$ 
and adding axioms which can be used for translating theorems of ZF which can be 
expressed in $L$, into equivalent $L$-formulas. We then choose a \Godel numbering
of $L(\hbox{ZF'})$, for instance by extending the $\gamma$ numbering we have chosen for $L$. 
This new \Godel numbering of $\hbox{ZF'}$ gives us an $L$-formula $A(x,y)$ representing provability in $\hbox{ZF'}$. 
Without any loss of generality we may assume that 
all ``formal'' theories in which one may derive results which can be meaningfully translated into $L$, 
have already been dressed up this way. Note that if the original language of $T$ involves using
some of the the symbols of $L$ in different roles, then we may rename those symbols first.
If $T$ is a theory such that $T$ proves every theorem of PA, then we will say that 
$T$ {\it contains\/} PA.
 
In conclusion, as far as 
theories $T$ expressed in different languages and `logics' are concerned, the only 
thing we assume for $T$ to be covered by our formalism, is that it is possible to code theorems and proofs in 
the theory $T$ into natural numbers in such a way that provability of $L$-formulas 
corresponds to satisfaction of an $L$-formula $A(x,y)$ with the two free 
variables $x$ and $y$, as described above. This assumption clearly holds for all formal theories $T$ one might use for 
deriving sound facts about $\nat$, since such theories should be recursively axiomatizable. 

On the other hand, of course, far from all $L$-formulas $A(x,y)$ with the free variables $x$ and $y$ can be interpreted meaningfully as representing provability of $L$-formulas in an actual ``theory''. 

Now let $G$ be the set of $\gamma$ numbers of $L$-formulas $A(x,y)$ with the free variables $x$ and $y$.
If the $L$-formula $A(x,y)$ corresponds to provability of $L$-formulas in a theory $T$ as in (1), and $g=\lceil A(x,y)\rceil$, then we say that $g\in G$ {\it represents\/} the theory $T$. Then also, if $B$ is an $L$-formula such that $T$ proves $B$, so that 
$\exists y A(\lceil B\rceil,y)$ is true, we say that the natural number $g=\lceil A(x,y)\rceil$ {\it proves\/} the formula $B$.  

Henceforth, by a {\it formal theory\/} we mean a theory represented by some $g\in G$. Note that under this definition, a
formal theory will be represented by infinitely many different numbers $g\in G$, corresponding to logically equivalent
formulas $A(x,y)$. A number $g\in G$ will be called {\it arithmetically sound\/}, or just {\it sound\/}, if 
all $L$-sentences proved by $g$ are true. Also, a theory will be called {\it sound\/} if it is represented by some sound $g\in G$.
 
Let $\Gamma$ be the set of $\gamma$ numbers of PA proofs of $L$-sentences on the form
$$
\exists y A(\lceil {\bf 0}={\bf 1}\rceil,y),
$$
where $\lceil A(x,y)\rceil\in G$. Then $\Gamma$ contains the set of $\gamma$ numbers of PA
inconsistency proofs for theories represented in $G$. The elements $p\in\Gamma$ will be called {\it lottery numbers}. 
Order $\Gamma$ by increasing size $\Gamma=\lbrace p_1,p_2,p_3,\ldots\rbrace$, so that by definition  
$p_n$ denotes the $n$th element in $\Gamma$ for $n\in\nat^+$, where $\nat^+=\nat\setminus\lbrace 0\rbrace$. 

\vskip.5cm

\proclaim{Definition 1}
{Let $g=\lceil A(x,y)\rceil\in G$. If $p\in \Gamma$ is the $\gamma$ number of a PA proof for \ 
$
\exists y A(\lceil {\bf 0}={\bf 1}\rceil,y)
$,
then then we say that 
$p\in\Gamma$ is a {\it lottery number for $g$}. If $g$ has a lottery number, then $g$ is called {\it inconsistent\/}. Otherwise, $g$ is called {\it consistent\/}.}

\vskip.5cm

\noindent If provability in a theory $T$ may be expressed by a formula $A(x,y)$ as in (1), and $p$ is a lottery number for 
$g=\lceil A(x,y)\rceil$, then we also say that $p$ is a {\it lottery number for the theory $T$\/}. So a
lottery number for $T$ is the $\gamma$ number of a PA proof for an $L$-sentence stating that the $L$-formula ${\bf 0}={\bf 1}$ is provable in $T$. In other words, a lottery number for $T$ is the $\gamma$ number of a PA inconsistency proof of $T$.

For defining our family of decision problems, we will use a diagonal construction linking each proof $p_n$ to two digit
sequences of lenght $n$. This can be done in many ways. Here we will use the infinite sequence $\varphi$ of base 10 digits obtained by starting with the set  of 1-digit sequences, followed by the set of 2-digit sequences and so on, each set ordered lexicographically. Then $\varphi=0123456789000102030405060708091011...$ in shorthand notation.
For $m,n\in\nat^+$, let $\varphi[m]_{[n]} $
denote the group of $n$ digits in $\varphi$ starting with the $m$th 
element. Thus we have 
$\varphi[15]_{[3]}=020$ and $\varphi[1]_{[2]}=01$. Moreover, if $k\in\nat$, then by the notation 
$t_k(\varphi[m]_{[n]})$
we mean $\varphi[m]_{[n]}+k$ modulo $10^n$ read as an $n$-digit natural number, possibly with initial digits zero. So this is the number obtained from $\varphi[m]_{[n]} $ by reading it as a natural number, adding $k$, taking the result modulo $10^n$,
and if needed, adding zeros in front to preserve $n$ digits. For instance $t_{97}(\varphi[15]_{[3]})=t_{97}(020)=117$
and $t_0(\varphi[15]_{[3]})=\varphi[15]_{[3]}=020$. 

For defining one of the two sequences to be associated with $p_n\in\Gamma$, we will use $p_n$ directly. For the other one, in order to obtain a sufficient difference between the two, a convenient option for our purposes is to use the natural exponential function, followed by the floor function picking out the integer part of function values. We will let $\exp:\nat\to\nat$ denote that composite function. Thus, e.g., $\exp(1)=2$ and $\exp(3)=20$.

\vskip.5cm

\proclaim{Definition 2}
{Let $k\in\nat$ and $n\in\nat^+$. If
$$
\varphi[\exp(p_n)]_{[n]}=t_k(\varphi[p_n]_{[n]}),\eqno(2)
$$
then $n$ will be called a {\it winner\/} in the $k$-lottery. The corresponding number $p_n\in\Gamma$ is then called a {\it winning 
lottery number in the $k$-lottery\/}.}

\vskip.7cm

\noindent We are now ready to describe our family $D=\lbrace D_k\mid k\in\nat\rbrace$ of decision problems. 

\vskip.5cm

\proclaim{Definition 3}
{Let $k\in\nat$. We define the decision problem $D_k$ as follows: Given $g\in G$, decide if there exists a winning lottery number $p\in\Gamma$ for $g$ in the $k$-lottery.}

\vskip.8cm 

\noindent For each $k\in\nat$, let  FW$_k$ be the $L$-sentence
$$
\exists m\forall n (n>m\longrightarrow \varphi[\exp(p_n)]_{[n]}\neq t_{\bf k}(\varphi[p_n]_{[n]})\hskip.1cm).
$$
Then  FW$_k$ states the existence of an upper bound $m$ on winners $n\in\nat$ 
in the $k$-lottery. Or, if you will, it states that there are a finite number of winners in the $k$-lottery. Hence the FW$_k$ designation. 

\vskip-.2cm

\section{3. Basic results} 

\vskip.3cm

\proclaim{Theorem 1} {PA proves $\forall k(\hbox{FW}_k\rightarrow  D_k\in\hbox{NP}).$}  

\vskip.5cm 

\proof We reason in PA. Let $k\in\nat$, and assume  FW$_k$. 
We will construct an algorithm $M_k$ which may be verified as being a polynomial time certificate checking algorithm for the decision problem $D_k$. To this end, let $M_k$ be the following algorithm: 

\vskip.3cm

\item{(i)} Given as input $g\in G$ along with a proposed certificate $p$ for $g$ relative to the decision problem $D_k$, check if $p$ is the $\gamma$ number of a correct PA proof of $\exists y A(\lceil {\bf 0}={\bf 1}\rceil,y)$, where 
$\lceil A(x,y)\rceil=g$. If not, then halt with answer ``Reject''. 

\vskip.2cm

\item{(ii)} It is now known that $g\in G$ and $p\in \Gamma$. By brute force checking of 
the $\gamma$ numbers up to $p$, determine $p_1$, $p_2$ and so on, until the $n$ such that $p_n=p$ is reached. Since the algorithm did not halt in step (i), such an $n$ will eventually be found. 

\vskip.2cm

\item{(iii)} Check if $\varphi[\exp(p_n)]_{[n]}=t_k(\varphi[p_n]_{[n]})$.
If so, halt with answer ``Accept''. Otherwise, halt with answer ``Reject''.

\vskip.3cm

\noindent Note first that stage (i) clearly can be performed by a Turing machine halting on all inputs, in polynomial time relative to the size of $p$, and therefore
clearly also in polynomial time relative to the size of $(g,p)$. 
\vskip0cm
Concerning stages (ii) and (iii), by FW$_k$ there exists an upper bound $m\in\nat$ on winners in the $k$-lottery. Thus a winning $p_n\in \Gamma$ must have $n\leq m$. So if 
the proposed certificate $p$ is correct, then the number of  computational steps required for stages (ii)--(iii) will be upwards bounded by a constant polynomial $\mu$, where the natural number $\mu$ is the maximum number 
$\mu$ of steps needed for performing these stages for an element $p_n\in \Gamma$ with index $n\leq m$. In particular, this number of steps is polynomial in $g$.
\vskip0cm
For an {\it incorrect\/} proposed certificate, the maximum number of computational steps spent on stages (ii)--(iii) will clearly be polynomial in the size of $(g,p)$. The reason is that if the algorithm actually reaches stage (ii) before halting, 
then only the $\gamma$ numbers up to $p$ needs to be checked in order to determine the $n$ such that $p=p_n$.  
\vskip0cm
The line of reasoning in PA described in the previous paragraphs, essentially proves the theorem. However, since the formal definition of NP plays an important role here, we need to be extra careful at this point. 
To check in full detail that 
the algorithm $M_k$ explicitly constructed above really can be verified in PA as a certificate checking algorithm for $D_k$
when assuming  FW$_k$, in the sense required by the ``official'' definition of NP given in \citeclay, we translate into the notation used there. Let 
$$
\Sigma=\lbrace 0,1,\ldots,9\rbrace
$$ 
be the alphabet consisting of the digits 0 to 9, and define the language $L_k$ as the set of strings $w\in \Sigma^\ast$ such that $w\in G$ and $w$ has a winning lottery number in the $k$-lottery. Define the binary relation 
$$
R(w,p)
$$ 
on $\Sigma^\ast\times\Sigma^\ast$ 
by letting $(w,p)\in R$ iff $w\in G$ and $p$ is winning lottery number for $w$, and let
$L_R$ be the language over 
$
\Sigma\cup\lbrace\#\rbrace
$ 
defined by
$$
L_R=\lbrace w\#p\mid R(w,p)\rbrace.
$$
We noted above that the algorithm $M_k$ is polynomial time in $(g,p)$. Translating,
$M_k$ is polynomial time in $w\# p$. Hence $L_R\in\hbox {P}$, 
with $M_k$ being a witness algorithm. Now since by FW$_k$ there are
a finite number of winners in the $k$-lottery, there exists an integer $j$ such that all
winning lottery numbers $p\in\Sigma^\ast$ have length 
$$
\vert p\vert<2^j.
$$
Then for all strings $w\in\Sigma^\ast$ we have
$$
w\in L_R\Longleftrightarrow 
\exists p(\hbox{$\vert p\vert<\vert w\vert^j$ and $R(w,p)$}),
$$
since clearly the length $\vert w\vert$ of any $w\in G$ with a winning lottery number satisfies $\vert w\vert\geq2$.
It follows that $R$ is a polynomial-time checking relation for $L_k$ in the terminology of \citeclay.  
Hence 
$L_k\in\hbox{NP}$.
So $D_k\in\hbox{NP}$, with $M_k$ 
being a certificate checking algorithm for $D_k$. This completes our reasoning in PA, and the theorem follows.\endproof

\vskip.6cm

\proclaim{Theorem 2}{Let $g\in G$ be consistent and containing PA. Then $g$ does not prove 
$$
\forall k(\hbox{${\bf g}$ has no winning lottery number in the $k$-lottery.})
$$}

\proof Reason in PA. Assume that $g\in G$ is inconsistent. Then there is a lottery number $p_n\in\Gamma$ for $g$. Now let 
$k=\varphi[\exp(p_n)]_{[n]}-\varphi[p_n]_{[n]}$ modulo $10^n$. 
Then $p_n$ is a winning lottery number for $g$ in the $k$-lottery. 
\vskip0cm
Combining the above PA reasoning with \Godels second incompleteness theorem [2] translated in the obvious way and applied to $g$, the theorem follows.\endproof

\vskip.7cm

\proclaim{Theorem 3} 
{Let $k\in\nat$, and let $T$ be a sound theory represented in $G$, containing PA.
Assume that there exists an explicitly constructed algorithm $M_k$ verified in $T$ as a solution to $D_k$. Let $g\in G$ be consistent. Then $T$ proves the following:
\vskip.2cm
For all $n\in\nat$, if $p_n\in\Gamma$ is a lottery number for ${\bf g}$, then
$$
\varphi[\exp(p_n)]_{[n]}\neq t_{\bf k}(\varphi[p_n]_{[n]}).
$$}

\proof Assume that there exists an explicitly constructed algorithm $M_k$ verified in $T$ as a solution to
the decision problem $D_k$. Since $T$ is sound, it follows that $M_k$ actually solves $D_k$ correctly. On the other hand, since $T$ contains PA, either

\vskip.1cm

\item{(a)}$T$ proves that $M_k({\bf g})$ halts with answer ``yes'', \ or

\item{(b)}$T$ proves that $M_k({\bf g})$ halts with answer ``no''.

\vskip.1cm

\noindent If (a), then since $T$ verifies $M_k$ as solving $D_k$ correctly, $T$ proves inconsistency of $g$. This contradicts soundness of $T$. Hence (b) holds, and the conclusion of Theorem 3 follows.\endproof

\section{4. The CC and FC conditions} 

\vskip.3cm

\noindent
In accordance with the discussion in Section 1, by a {\it constructible algorithm\/} we simply mean
an algorithm we are able to construct explicitly in our real world. Hence we consider the concept of
``constructible algorithm'' as external to formal theories. 
Nevertheless, as mentioned in Section 1, no algorithm is useful without a verifying proof in some theory which we can actually construct, that is, a constructible theory.  

Of course, there is no precise definition of ``constructible theory'' exactly representing the full scope of the informal real-life concept. However, in order to proceed we only need to define a precise {\it necessary\/} condition on constructible theories which is strict enough to make our proofs go through, while being weak enough for it to arguably be satisfied by all theories which are real-life constructible. Since we need only to deal with theories $T$ which are {\it sound and represented in $G$\/}, we limit ourselves to this set of theories. The constructibility condition we will
assume is descibed in the statement CC given below.

\vskip.5cm

\item{{\bf CC}}
Let $T$ be a sound, constructible theory containing PA, represented by $g\in G$.  If there exists $k\in\nat$ such that $T$ proves
$$
\neg\exists n(\hbox{($p_n$ is a lottery number for ${\bf g}$)}
\wedge \varphi[\exp(p_n)]_{[n]}=t_{\bf k}(\varphi[p_n]_{[n]}),\eqno(3)
$$
then $T$ also proves
$$
\neg\exists k\exists n(\hbox{($p_n$ is a lottery number for ${\bf g}$)}
\wedge \varphi[\exp(p_n)]_{[n]}=t_k(\varphi[p_n]_{[n]})).\eqno(4)
$$
In other words: If there is $k\in\nat$ such that $T$ proves the nonexistence
of a winning lottery number for $T$ in the $k$-lottery, then $T$ also proves that there is no winning number for $T$ in any of the lotteries corresponding to the decision problems in $D$.

\vskip.5cm

\noindent To justify CC, let $T$ be a sound, constructible theory represented in $G$ containing PA, and let $g\in G$, $k\in\nat$. Assume that $T$ proves (3).

While the $\gamma$ number $p_n\in\Gamma$ corresponding to a given index $n$  can be calculated in any sound theory containing PA, in each proof such a brute force method can only be used for a finite number of indices $n$. Hence in each $T$-proof, the digit groups $\varphi[p_n]_{[n]}$ and $\varphi[\exp(p_n)]_{[n]}$ corresponding to a given index $n$ can only be calculated for a finite number of $n\in\nat$. Thus if $T$ proves (3), there exists $m\in\nat$ such that $T$ proves (3) 
without calculating neither $\varphi[p_n]_{[n]}$ nor $\varphi[\exp(p_n)]_{[n]}$ for any $n>m$. This is certainly possible 
in cases where $T$ proves (4), for instance it will happen if $T$ proves consistency of $g$. What we want to show, is that if there exists $m\in\nat$ such that 
$T$ proves (3) {\it without\/} calculating 
$\varphi[p_n]_{[n]}$ or $\varphi[\exp(p_n)]_{[n]}$ for any $n>m$, the $T$ also proves (4). 
\vskip0cm
The key here is that if for all sufficiently large $n\in\nat$ the digit groups 
$\varphi[p_n]_{[n]}$ and $\varphi[\exp(p_n)]_{[n]}$ are not used in the proof of (3), then there will be no use for $k$ either. The reason is that $k$ enters into (3) only through the translation $t_k$, which provides no additional restrictions or symmetry breaking for indices $n$ whose 
corresponding $\varphi[p_n]_{[n]}$ and $\varphi[\exp(p_n)]_{[n]}$ sequences are unknown. Concerning this symmetry, note that the sequence  $\varphi$ is {\it asymptotically normal\/} in the sense that all 1-digit sequences occurs equally often in $\varphi$ from the beginning of $\varphi$, all two-digit sequences occur equally often from the group of 2-digit sequences in $\varphi$  onwards, and in general all $n$-digit sequences occur equally often in $\varphi$ starting from the group of $n$-digit sequences in $\varphi$ onwards.
More precisely, for all $k\geq n$ and
$m$ in the range $10^{k-1}\leq m<10^k$, all $n$-digit sequences appear as $\varphi[m]_{[n]}$ the same number of times.
Note that since $\exp(p_n)$ will always have more than $n$ digits, this normality condition always applies to (3). 
\vskip0cm
On the other hand, by \Godels second incompleteness theorem [2], there will be an infinite number of consistent theories represented in $G$ for which $T$ cannot prove consistency. For any one of these theories, a single,
additional, unknown inconsistency proof $p_a$ would offset the relation
$\varphi[\exp(p_n)]_{[n]}=t_{\bf k}(\varphi[p_n]_{[n]})$ for all
$n>a$, in that the new digit group $\varphi[p_n]_{[n]}$ would 
be what was originally
$\varphi[p_{n-1}]_{[n]}$, while the new $\varphi[\exp(p_n)]_{[n]}$ would 
be what was originally
$\varphi[\exp(p_{n-1})]_{[n]}$. This shows that any {\it rational\/} argument used in $T$ for proving (3) by somehow linking the value
of $k$ to the ordering of $\Gamma$, would need to work even under arbitarily large and unknown (and actually nonexistent!) offsets of this kind. Assuming this is impossible in a constructible and sound theory, we conclude that any proof of (3) in $T$ must work independently of $k$, and hence 
be generalizible to a proof of (4) in $T$.

Finally, there is the possibility that (3) follows directly from an axiom of $T$, so that no rational reasoning concerning the ordering of $\Gamma$ whatsoever is involved in deriving (3).  A trivial case of this would be that (3) itself is an axiom of $T$.
 
Here the constructibility assumption on $T$ again comes into play. Combined with the soundness assumption on $T$, constructibility requires that the axioms of $T$ are known (believed) by us to be sound. Hence if (3) is an axiom of $T$, we must have verified its soundness by some previous argument. If the theory $T'$ we used for that verification is also constructible, which it should be in order to validate soundness of $T$, then we may shift our focus from $T$ to $T'$. There can be no infinite regress $T, T', T'',...$ here, since our soundness validitation process must have started from a theory considered as sound {\it a priori\/}. In other words, without loss of generality we may assume that the proof of
(3) in $T$ represents a rational line of reasoning somehow involving the ordering of $\Gamma$. This is the situation discussed previously, and our justification of CC is complete.\endproof

\vskip.7cm

\noindent We now turn to the aforementioned finiteness condition. This is the following:

\vskip.5cm

\item{{\bf FC}}There exists $k\in\nat$ such that FW$_{k}$ is true.

\vskip.3cm

\noindent While the validity of FC may be considered ``obvious'', and it can be argued for in many ways, it is in fact questionable whether or not FC is provable in any constructible, sound theory at all. Here we touch onto the potential problem of {\it absolute undecidability\/} in mathematics. We will return to this in the
final section of the paper. 

As our justification of FC, we will use an idea which may be interpreted geometrically by imagining the $k$-lotteries as an infinite row of nests numbered $k=0,1,2,3\ldots$. The nests receive eggs (``winners'') in egg-laying rounds $n=1,2,3,\ldots$, with each nest receiving either one egg or no eggs in each round. In round $n$, nest $k$ receives an egg iff 
$
\varphi[\exp(p_n)]_{[n]}=t_k(\varphi[p_n]_{[n]})
$. 
Thus in round $n=1$ of egg-laying, nests with $k$ such that $\varphi[\exp(p_1)]_{[1]}=t_k(\varphi[p_1]_{[1]})$ receives an egg. In other words, every tenth $k$-nest recieves an egg, starting with some $k$ such that $0\leq k\leq 9$. Thus the contribution to the average number of eggs across all $k\in\nat$ from round $n=1$ og egg-laying, 
asymptotically approaches $1/10=10^{-1}$. 
\vskip0cm
For $n=2$, the pattern repeats. Since there there are $10^2=100$ different 2-digit combinations, every 100th $k$-nest receives an egg, starting with 
some $k$ such that $0\leq k\leq 99$. The contribution to the asymptotic average number of eggs is $1/100=10^{-2}$. In general, in the $n$th round of ``egg laying'', every $10^n$th $k$-nest receives an egg, with the first one satisfying $0\leq k\leq 10^n -1$. The contribution to the asymptotic average number of eggs is $10^{-n}$. What we need to show in order to justify FC, is that in the limit when 
all egg-laying rounds $n\in\nat^+$ are completed, then at least one nest will have received a finite number of eggs only. 
\vskip0cm
To see why this conclusion is natural, let $\overline{w}_{0\leq k\leq M}^N$ be the average number of eggs in nest $k$ for $k=0,\ldots,M$ at the point where eggs from rounds $n=1,\ldots,N$ are completed. That is, $\overline{w}_{0\leq k\leq M}^N$ is the total number of eggs in nests $k$ for $k=0,\ldots,M$ at this point, divided by $M+1$. We have
$$
\lim_{N\to\infty}\Big[\lim_{M\to\infty}\overline{w}_{0\leq k\leq M}^N\Big]
\leq\lim_{N\to\infty}\Big[\sum_{n=1}^N 10^{-n}\Big]={1\over 9}.
$$
Based on this, taking into account the asymptotically normal property of $\varphi$
described in the justifiction of CC,
it is resonable to assume that the average number of eggs received in each $k$-nest when all the egg-laying rounds 
$n\in\nat^+$ has been completed, is $1/9$. Translating into probabilistic language, in the canonical random model the expected total number of eggs in each nest is $1/9$. Moreover, the random model probability for the event that a given, fixed $k$-nest receives an infinite number of eggs, is zero. 
\vskip0cm
While a great deal more can be said about this, in this paper we will simply 
take the above remarks as our motivation for assuming that there must be {\it at 
least one\/} nest $k_f$ receiving a finite number of eggs only. (In fact, it is reasonable to believe that there will be infinite number of such nests, or even that no nest will receive an infinite number of eggs.) Under this assumption FW$_{k_f}$ is true, and FC is justified.\endproof

\vskip.8cm

\proclaim{Theorem 4}
{Assume CC, and let $k\in\nat$ be arbitrary and fixed. Then there is no sound, constructible theory represented in $G$ verifying a solution algorithm $M_k$ for $D_k$.}

\vskip.4cm

\proof Let $T$ be sound, constructible and represented by $g\in G$. Without loss of generality we may assume that $T$ contains PA. Given $k\in\nat$, assume for contradiction that $g$ verifies an algorithm $M_k$ solving $D_k$. By Theorem 3 it follows that
$g$ proves the nonexistence of a winning lottery number for $g$ in the $k$-lottery. By
CC, this implies that $g$ also proves
$$
\forall k(\hbox{${\bf g}$ has no winning lottery number in the $k$-lottery}).
$$ 
But since $T$ is sound and hence consistent, by Theorem 2 this is impossible.\endproof

\vskip.8cm 

\proclaim{Theorem 5}
{Assume CC and FC. Then there exists $k\in\nat$ such that (i) $D_k\in\hbox{NP}$, and (ii) No sound, constructible theory represented in $G$ verifies a solution algorithm for $D_k$.}

\vskip.4cm

\proof By FC, let $k\in\nat$ be such that FW$_k$ is true. Then by Theorem 1, using soundness of PA, 
$$
D_k\in\hbox{NP}.
$$
On the other hand, combining CC with Theorem 4, it follows that no sound, constructible theory represented in $G$ verifies a solution algorithm for 
$D_k$.\endproof

\eject

\section{5. Discussion}

\noindent At this point, the reader may feel that clarification of a particular point is urgent: In Theorem 5, one may replace P by any class consisting of algorithmically solvable decision problems, for instance EXPTIME.
How is that possible, given that there are well known proofs of NP $\subseteq$ EXPTIME?  

The reason for this is that all classical proofs of
NP $\subseteq$ EXPTIME tacitly use a specific assumption 
which 
normally is justified, but which may not hold for 
the decision problems $D_k$ in the family $D$. However, this issue is not as serious as one might think at first glance. The classical proofs may be fixed by adding an assumption which is easily verifiable in ``normal'' cases, preserving the conclusion NP $\subseteq$ EXPTIME in all common theoretical and practical applications.

To see which additional assumption is needed here, note that the 
certificate verifying algorithm $M_k$ for $D_k$ constructed 
while assuming FW$_{k}$ in the proof of Theorem 1, does not involve the {\it value\/} of an upper bound on winning indices. Hence the algorithm $M_k$ {\it exists\/} in the constructible sense; it can be 
written down without knowing an upper bound on the set of winners. In the notation of \citeclay, used in the proof of Theorem 1, this corresponds to $M_k$ not employing the value of the integer $j$. In contrast, in order to modify $M_k$ into an algorithm 
{\it solving\/} the problem $D_k$, one would need the numerical value of of an upper bound on winners, or 
on $j$ if you will. However, such upper bounds may be impossible to derive.

On the other hand, if we assumed that an explicit value of an upper bound $m$ as in the proof of Theorem 1 was available in all cases where $L_R\in\hbox{P}$, then we could describe NP simply by saying that a language is in NP iff a proposed certificate
for a given string $w$ in the language can be correctly decided (either accepted or rejected) by an algorithm 
in polynomial time relative to the length of $w$. More or less precise descriptions of NP similar to this seem to be widespread, but they all tacitly assume that the explicit value of an upper bound on $j$ can be found and subsequently used in constructing a solution algorithm for the problem. 

It is illustrative to compare the above considerations to a concrete NP $\subseteq$ EXPTIME proof. Consider the following example: 

\vskip.2cm

\item{}``If a language is in NP, then each string in the language has a certificate whose size is polynomially bounded in the size of the string. Moreover, each of these certificates may be verified in a number of steps polynomially bounded in the length of the string. Thus with a finite alphabet, there is an exponential time solution algorithm which checks all certificates of length up to the polynomial length bound determined by a given input string. So the language is in EXPTIME.''

\vskip.2cm

\noindent When applied to the decision problems $D_k$, the phrase ``there is'' in the above proof becomes problematic.
In order to construct an exponential time 
solution algorithm for a decision problem based on a polynomial time certificate verifier, using the brute force checking idea, we need an
{\it explicitly given\/} polynomial bounding the size of correct certificates. In the case of $D_k$, there is no reason to believe that such an explicitly constructible polynomial is available. Hence the above proof of NP $\subseteq$ EXPTIME breaks down. However, if we restrict the proof to languages in NP for which a polynomial bounding the size of correct certificates for $w$ 
in terms of $\vert w\vert$ may be explicitly constructed, then the proof regains its validity. In cases of practical interest, this condition will usually be satisfied.

Turning now to the big question: Does Theorem 5 solve the P versus NP problem, showing that $\hbox{NP}\not\subseteq\hbox{P}$ under the constructive interpretation of complexity classes? 

We will describe two reasons for answering ``yes''. First, there are the justifications we have given for CC and FC. Despite not being formalizable, these may be considered obviously valid from a more general perspective. Accepting them, then under the constructive interpretation of NP and P, Theorem 5 gives us the existence of an integer $k$ such that 
$D_k\in\hbox{NP}$ and $D_k\notin P$. Hence $\hbox{NP}\not\subseteq\hbox{P}$.

Secondly, concerning the possibility of ``doing better'' by finding a proof of $\hbox{NP}\not\subseteq\hbox{P}$ which is 
formalizable in a particular constructible theory assumed to be sound, such as PA or ZF, there are well known counter\-arguments. In fact, it may be considered somewhat surprising that these arguments, which are rather simple and striking, 
seemingly have not received more attention. 

To decribe what may be considered the main counterargument, expressed in our formalism, it is convenient to use the original {\it nondeterministic polynomial\/} definition of NP, equivalent to the certificate checking definition commonly used today [1]. The argument is very simple: If there is a nondeterministic algorithm solving a given decision problem $\Delta$ in polynomial time,
then working in a fixed, constructible and sound theory $T$ represented in $G$, it is impossible to theoretically eliminate the
possibility that there might exist a deterministic algorithm solving the problem equally fast. Theoretically eliminating this possibility would require finding a {\it theory-independent
limitation\/} on how ``smart'' a polynomial time algorithm can be. That is, $T$ would need to establish a theory-independent upper limit on the 
resolving power of {\it theorems from more powerful theories\/} which a polynomial time algorithm might use for finding the solution in some ``trivial'' way; a way which viewed from the outside of the more powerful theory could look just as if the algorithm makes
a sequence of correct guesses based on theorems which are undecidable in $T$, enabling it to construct a solution in the same computational time as it would take for a nondeterminstic algorithm to solve the problem. By the assumption
$\Delta\in\hbox{NP}$, this latter 
computational time is polynomial. 

Deriving such a theory-independent limitation inside a single constructible and sound theory $T$ represented in $G$, can be considered as {\it obviously\/} impossible. By \Godels first incompleteness theorem [2], $T$ cannot characterize any theory-independent limitation on the set of $\nat$-true sentences of $L$ derivable in sound, formal theories.

\vskip-.1cm

\section{6. Absolute undecidability}

\noindent The construction employed in this paper leads to effects which may be considered as representing a form of {\it absolute undecidability\/}. For instance, we have the following result.

\vskip.5cm

\proclaim{Theorem 6}{Assume CC and FC. Let $\kappa$ be the smallest $k\in\nat^+$ such that FW$_k$ is true, let $m\in\nat$
be the smallest integer such that all winners $n$ under $\kappa$ satisfy $n<m$, let $\hbox{Prime}_m$ be the $m$th prime number, and let $\xi=(\hbox{Prime}_m)^\kappa$. Then no sound, constructible theory represented in $G$ decides the value of 
the natural number $\xi$.}

\vskip.4cm

\proof Let $T$ be sound and constructible, and assume that $T$ decides $\xi$. Without loss of generality, we may assume that $T$ contains PA. Then $T$ also decides $\kappa$ and $m$. By brute force
checking of $p_n$ for $n$ up to $m$, $T$ may then decide if it has a winning lottery number in the $\kappa$-lottery. Since $T$ is sound, the answer will be no. Then by CC, $T$ proves that it has no winning lottery number in any $k$-lottery. By theorem 2, this is impossible.
\endproof

\vskip.5cm

\noindent Together, CC and FC thus imply the existence of a natural number $\xi$ which is uniquely definable in the language of PA, but which is {\it absolutely undecidable\/} in the sense that it cannot be calculated in any sound, constructible theory represented in $G$. This effect arguably has consequences going far beyond the P versus NP problem. However, we will not go deeper into these matters here.

\vskip.7cm

\centerline{{\bf Bibliography}}

\vskip.2cm

\item{[1]} Cook, S. {\it The P versus NP Problem.\/} Clay Mathematics Institute, 2(6), 2000.

\vskip.2cm

\item{[2]} \godel, K. {\it Collected Works Vol I.\/} Oxford University Press, 1986.

\bye